\documentclass{aa} 

\usepackage{lineno}
\usepackage{graphicx}
\usepackage{upgreek}  
\usepackage{siunitx} 
\usepackage{xcolor}
\usepackage{marginnote} 
\begin{document}

   \title{Interpretation through experimental simulations of phase functions 
   revealed by Rosetta in 67P dust coma} 

  \titlerunning{Experimental analogs of the phase functions of 67P}

   \author{A.C. Levasseur-Regourd\inst{1}
          \and
          J.-B. Renard\inst{2}
          \and
          E. Hadamcik\inst{3}
	\and
          J. Lasue\inst{4}
	\and
          I. Bertini\inst{5}
	\and
          M. Fulle\inst{6}
          }
 
   \institute{LATMOS, Sorbonne Univ., CNRS, UVSQ, Campus Pierre et Marie Curie, 4 place Jussieu, 75005 Paris, France\\
             \email{aclr@latmos.ipsl.fr}
         \and
            LPC2E-CNRS, 3A avenue de la Recherche Scientifique, Orléans cedex 2, France
         \and 
            LATMOS/IPSL, UVSQ Université Paris-Saclay, Sorbonne Université, CNRS, Guyancourt, France
         \and
            IRAP, Université de Toulouse, CNRS, CNES, UPS, 9 avenue Colonel Roche, 31400 Toulouse, France
         \and
            Department of Physics and Astronomy ‘G. Galilei’, University of Padova,
Vic. Osservatorio 3, I-35122 Padova, Italy
         \and
            INAF - Osservatorio Astronomico, Via Tiepolo 11, I-34143 Trieste, Italy
             }

   \date{Received September 15, 1996; accepted March 16, 1997}
 
  
  \abstract
   {The dust-brightness phase curves that have been measured by 
   the OSIRIS cameras on board the Rosetta spacecraft within the 
   coma of comet 67P/Churyumov-Gerasimenko (67P) present a remarkable u-shape. }
   {Our goal is to compare these phase curves with those of tentatively 
   analog dust samples to assess the key dust properties 
   that might induce this shape.}
   {Light-scattering measurements have been made with the PROGRA2 
   instrument in the laboratory and in microgravity conditions 
   on samples of different physical properties and compositions that are 
   likely to be representative of cometary dust particles.}
   {We find that the brightness phase 
   curves of a series of interplanetary dust analogs that have been recently 
   developed (to fit the polarimetric properties of the inner zodiacal 
   cloud and their changes with heliocentric distance) are quite comparable 
   to those of 67P.
   Key dust properties seem to be related to the composition and the porosity.}
   { We conclude that the shape of the brightness phase curves of 67P
     has to be related to the presence of a significant amount of organic
     compounds (at least 50\% in mass) 
   and of fluffy aggregates (of a size range of 10 to \SI{200}{\um}). We also confirm 
   similarities between the dust particles of this Jupiter-family comet 
   and the particles within the inner zodiacal cloud.}


   \maketitle
%

\section{Introduction}

  Everywhere in the solar system, dust media, such as cometary comae 
  or the interplanetary dust cloud, scatter solar light. While the properties 
  of the scattered light depend on the properties of the scattering medium, 
  it is a complex inverse problem to derive information from light-scattering
  measurements.
  The unique Rosetta rendezvous with comet 
  67P/Churyumov-Gerasimenko (thereafter 67P) has fortunately 
  provided a wealth of information on cometary dust particles during 
  its 26-month-long mission, including phase 
  functions of these particles \citep{bertini2017}.
  
  From past cometary flybys, including those of comet 1P/Halley in 1986, 
  to the 67P rendezvous, numerous observations of dust in cometary comae 
  have been obtained from the Earth and Earth-orbiting observatories.  
  On a given day, remote observations of a comet correspond to a given 
  phase angle $\alpha$ and to different distances to the nucleus within a 
  possibly heterogeneous coma. Retrieving a phase curve thus 
  requires extended observational periods, corresponding to significant 
  changes in solar distance and possible changes in activity during 
  outburst events \citep{dollfus1988}. Interpreting these phase curves 
  nevertheless provides information on the dust properties of various 
  comets, with realistic experimental and numerical simulations  now
  validated through 67P ground-truth. 
  
  \bigskip
  
  The Rosetta mission has provided a huge amount of data on dust
  particles and their phase curves over a 
  wide range of distances to the Sun and to the nucleus. Results stem 
  from innovative dust instruments, the COmetary Secondary Ion Mass
  Analyzer (COSIMA), the Grain Impact Analyzer and Dust Accumulator
  (GIADA), the Micro-Imaging Dust Analyzis System (MIDAS), as well as 
  from its Optical, Spectroscopic and Infrared Remote Imaging System
  (OSIRIS) camera and its Philae lander \citep{glassmeier2007}. 
  Rosetta dust studies, obtained on a wide range of sizes (at least from tens of 
  nanometers to a few meters) have clearly established what had   
  been suggested after Giotto and Stardust missions 
  \citep{fulle2000, horz2006, matrajt2008}: dust particles are aggregates
  that present a hierarchical 
  structure and have an average bulk porosity of at least 60\% 
  \citep{hilchenbach2016, bentley2016, langevin2016, 
  mannel2016}. The Rosetta mission has also provided outstanding 
  information on the composition of refractories in the coma, and it has shown 
  that complex organic molecules represent a significant component 
  of the solid fraction of comets 
  \citep{goesmann2015, fray2016, bardyn2017, fray2017}. 
  These results \citep[see, e.g., ][for a review]{levasseur-regourd2018}
  are of major interest for the interpretation of remote 
  observations of dust in the comae of comets, and especially of 
  Jupiter-family comets (JFCs) such as 67P. It may be added that while
  cometary dust and dust from asteroids 
  replenish the interplanetary dust cloud (also called zodiacal cloud), 
  JFCs have been understood based on various observational and 
  dynamical approaches to be the main sources of dust particles 
  along  the Earth's orbit 
  \citep{lasue2007, nesvorny2010, rowan-robinson2013, carrillo-sanchez2016}.


  In relation with optical properties of dust, OSIRIS multiwavelength 
  observations (from about \SI{376}{\nm} to \SI{744}{\nm}) have allowed 
  \citet{bertini2017} 
  to retrieve the shape of the coma-dust phase function between 
  about \ang{15} and \ang{150} 
  from data obtained between March 2015 and February 2016.
  The reflectance slowly decreases from about \ang{20} to \ang{100}
  before increasing again without any sharp surge in the forward-scattering region. 
  This flattened u-shape seems to agree with most observations of other
  comets \citep[][Fig. 5]{bertini2017} and rules out a strong forward-scattering.
  
  OSIRIS phase curves of the coma dust have a significantly flatter decrease 
  in backscattering than the nucleus phase curves \citep{fulle2018}. 
  They provide clues to a slight reddening for dust particles, lower than for the 
  nucleus, and to a negligible phase reddening, which is consistent with the absence of 
  multiple scattering in the coma \citep{bertini2017, fulle2018}. 
  It may finally be added that a phase curve 
  (from \ang{1.2} to \ang{74}) has also been derived from 
  OSIRIS data obtained between July 2014 and February 2015, 
  typically at \SI{612.5}{\nm} \citep{fink2018}. 
  It shows a fair agreement with the results described above in
  the \SIrange{15}{90}{\degree}  domain of 
  overlapping phase angles.
  
  \bigskip
  

  Some characteristics of the light scattered by the randomly polarized
  solar light on dust media (such as cometary comae or the 
  interplanetary dust cloud) may provide information on their properties.
  Characteristics of scattered light typically depend upon 
  the concentration, size, size distribution, shape, morphology, porosity, 
  and complex refractive index (related to the composition) of the dust
  particles, and thus to their 
  geometric albedo. Because these particles are mostly larger than 
  the observational wavelengths (usually in the visible and near-infrared 
  domains), the Mie theory cannot be used, except for almost spherical dust 
  particles or droplets.  
  
  Simulations are mandatory to try to unequivocally 
  interpret observational data in terms of the physical properties and possibly 
  the composition of the dust particles. Observations of the linear polarization 
  of the scattered light are somewhat easier to interpret than brightness 
  observations because they neither depend upon the distances of the scattering 
  medium to the Sun and to the observer nor upon the concentration of the medium. 
  For cometary comae of comets that have been extensively observed (1P/Halley and 
  C/1995 O1 Hale-Bopp), numerical and experimental simulations have 
  typically suggested that the dust particles are likely to be fluffy aggregates 
  of irregular grains mixed with some compact particles,
  and that they are composed 
  of minerals and absorbing carbonaceous material 
  \citep{hadamcik2007, levasseur-regourd2008, lasue2009}.
  Similar approaches have also been developed for 
  the interplanetary dust cloud \citep{lasue2007, hadamcik2018}. 
  
  
  In this paper, we first consider experimental simulations with analogs 
  that might be representative of cometary dust.
  The simulations have been mostly performed under 
  microgravity conditions with the PROGRA2 instruments.
  We then compare the phase curves of the analogs with those of dust
  particles in the coma of 67P 
  just before and soon after perihelion
  to point out satisfactory analogs. We then discuss our results 
  and their possible implications.

\section{PROGRA2 experimental simulations of light scattering with cometary analogs }

  Experimental simulations in the laboratory or under microgravity 
  conditions allow measurements to be made on various samples of controlled 
  characteristics. Reviews on technical developments for light-scattering 
  measurements in microgravity conditions, together with types  of 
  samples of interest in planetary sciences, and on experimental 
  scattering matrices of clouds of randomly oriented particles may be found in 
  \citet{levasseur-regourd2015} and in 
  \citet{munoz2015}, respectively. 
  We summarize below the instrumentation 
  we have used and the samples we have estimated to be of interest for 
  simulating cometary dust properties.

\subsection{PROGRA2 instrumentation}

  The PROGRA2-Vis instrument (PRopriétés Optiques des GRains Astronomiques 
  et Atmo\-sphé\-riques\--visible, i.e., Optical properties of astronomical and 
  atmospheric grains in the visible domain) is dedicated to the retrieval of 
  the brightness and linear polarization phase functions of levitating particles 
  with random orientations \citep{worms1999, renard2002}.
  The light sources at present are halogen white lamps with a depolarizing 
  filter and a spectral filter; one source operates at $555 \pm$\SI{30}{\nm},
  and the other at $650 \pm$\SI{30}{\nm}. An optical fiber carries the
  light to the vial 
  in which the particles are lifted. The particles that cross the light-beam 
  scatter the incident light. A polarizing beam-splitter cube splits 
  the scattered light into its two components, parallel and perpendicular 
  to the scattering plane. These are recorded by two synchronized 
  cameras with similar fields of view. The vial and a third synchronized 
  camera are mounted on a rotating device; the incident light beam and 
  the vial rotate to change the phase angle in 
  the \SIrange{8}{165}{\degree} range, 
  the detection system being in a fixed position. The third camera records 
  the scattered light at a constant phase angle of \ang{90}, and acts as a 
  reference camera. The polarization is retrieved from the two first cameras; 
  the brightness is retrieved after normalization of the flux recorded by the
  two first cameras to the flux of the third camera. In the following,
  we consider the brightness results  for further comparisons. 
  
  The levitation of the particles is obtained by two methods. 
  For compact particles smaller than about \SI{20}{\um} and for fluffy particles, 
  which are aggregates of submicron-sized grains (also called monomers), 
  the particles are lifted through an air draught technique 
  \citep{hadamcik2002}. 
  For compact particles larger than \SI{20}{\um} and for mixtures of particles 
  with different structures, the measurements are conducted in microgravity 
  conditions during parabolic flights on board the A300 ZeroG or the A310 ZeroG
  aircraft managed by the Novespace Company \citep{renard2002}.

\begin{figure*}
	\center
	\includegraphics[width = 6 in]{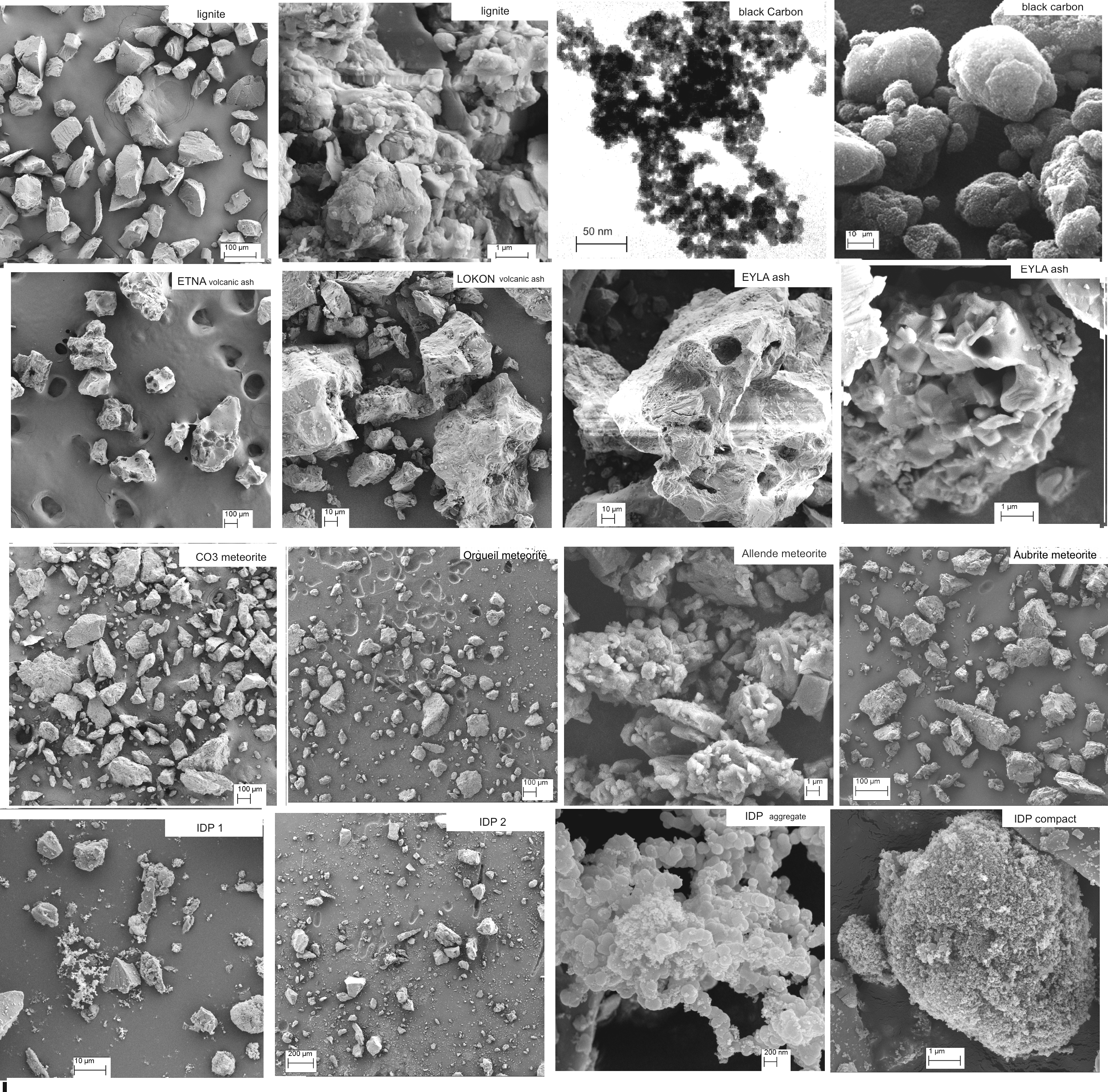}
        \caption{Typical SEM images of dust samples that might be representative
          of cometary dust properties in terms of size, morphology, and
          composition. From top to bottom, we present the black carbon
          fluffy aggregates, the porous volcanic ashes, the crushed
          meteoritic dust particles, and the interplanetary dust 
        analogs that we use for comparisons with OSIRIS phase functions. }
	\label{fig:fig1}
\end{figure*}

\subsection{Choice of samples}
  
  Four types of samples that could partly reproduce some physical properties 
  of cometary material have been chosen. Scanning electro microscope (SEM)
  images of these samples are shown on Fig.~\ref{fig:fig1}.
  
  
  Samples with morphologies that might be representative of cometary
  dust are considered 
  first. Black carbon fluffy aggregates, with a mean size of about 
  \SI{100}{\um}, of monomers of \SIlist{14;25;56;90}{\nm} may be used to 
  estimate the effect of the monomer sizes on the shape of the phase 
  function \citep[updated from]{francis2011}. Second, porous volcanic ashes 
  from Eyjafjallajökull (Iceland) and Lokon (Indonesia) with sizes
  below \SI{50}{\um} 
  and also from Etna (Italy), with three size-ranges of about 
  \SIlist{50;100;200}{\um}, 
  are considered as possibly representative of porous particles.
  
  Then, samples with compositions tentatively representative of cometary dust 
  are considered. Four samples of crushed meteoritic dust particles were
  obtained, 
  with sizes smaller than \SI{200}{\um}, one from an aubrite
  (i.e., achondrite meteorite) from 
  Antarctica, and, possibly more representative of the composition, carbonaceous 
  chondrites, such as Orgueil (CI1), Allende (CV3), or North Africa 6352 (CO3) 
  \citep{hadamcik2011}; 
  the size of the crushed particles is smaller than \SI{200}{\um}. 
  Finally, four possible interplanetary dust analogs, composed of mixtures of 
  carbonaceous and mineral compounds, with ratios that  change from one analog 
  to the other, were used. While each material exists in different structures, 
  the ratio between fluffy aggregates and more compact particles was
  kept constant in mass, 
  ($35 \pm 10$) \% for aggregates and ($65 \pm 10$) \% for compact particles. 
  While mass, volume, and count percentages certainly depend on the 
  various porosities of the particles, values of about 37\% in volume were 
  derived for extremely porous fractal particles in 67P coma by \citet{fulle2017}.
  Moreover, about 35\% in counts of type C tracks of dust particles collected by 
  Stardust within the coma of 81P/Wild 2 were found by \citet{burchell2008}. 
  Carbonaceous particles are fluffy carbon-black aggregates with submicron-sized 
  grains or porous coals such as lignite; mineral particles are
  silicates in fluffy aggregates 
  or in more compact particles, and some of them originate from crushed 
  meteorites \citep{hadamcik2018}. 
  The particles follow size distributions  of \SI{10}{\um} to \SI{320}{\um}.
  The percentage of carbonaceous content decreases 
  from 60\% to 30\% to tentatively reproduce the evolution with
  solar distance of the optical 
  properties of the cloud and thus of the particles composition 
  \citep{lasue2007}.

\section{Comparison between dust-coma phase functions of 67P and
    those obtained for analogs}
    
\begin{figure}
	\center
	\includegraphics[height = 2.2in]{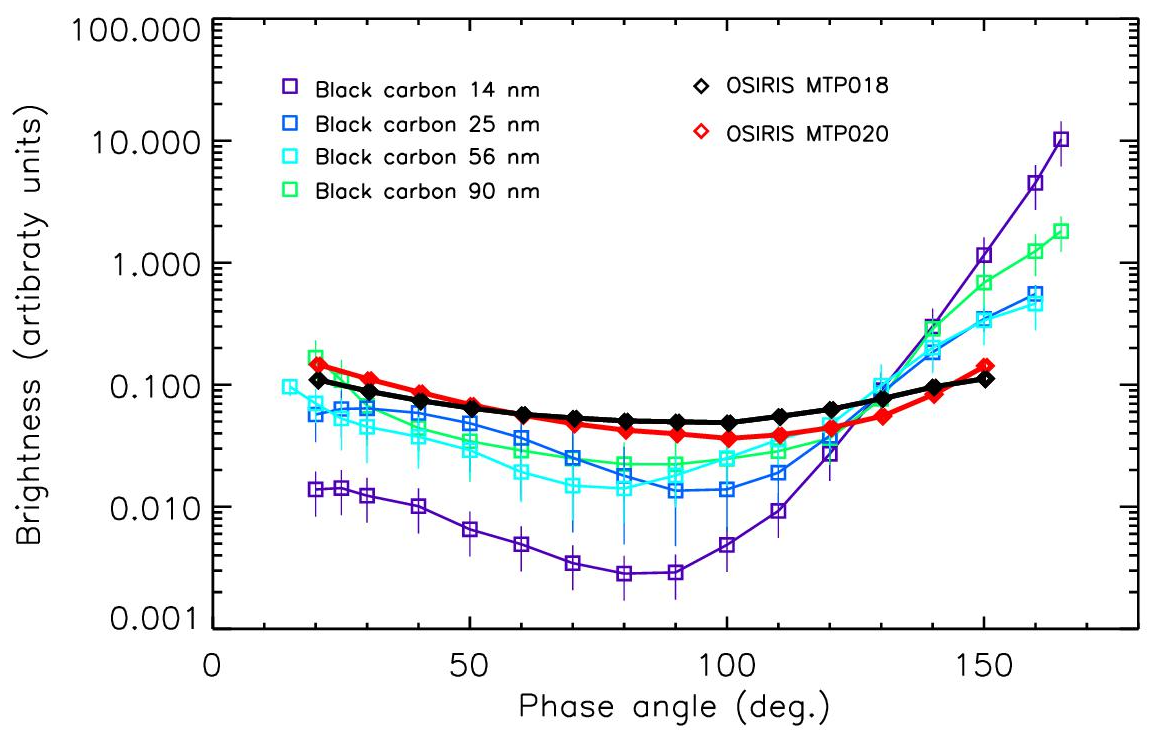}
	\caption{Comparison between OSIRIS and PROGRA2 phase functions
	for aggregates of black carbon particles with different sizes.}
	\label{fig:fig2}
\end{figure}

\begin{figure}
	\center
	\includegraphics[height = 2.2in]{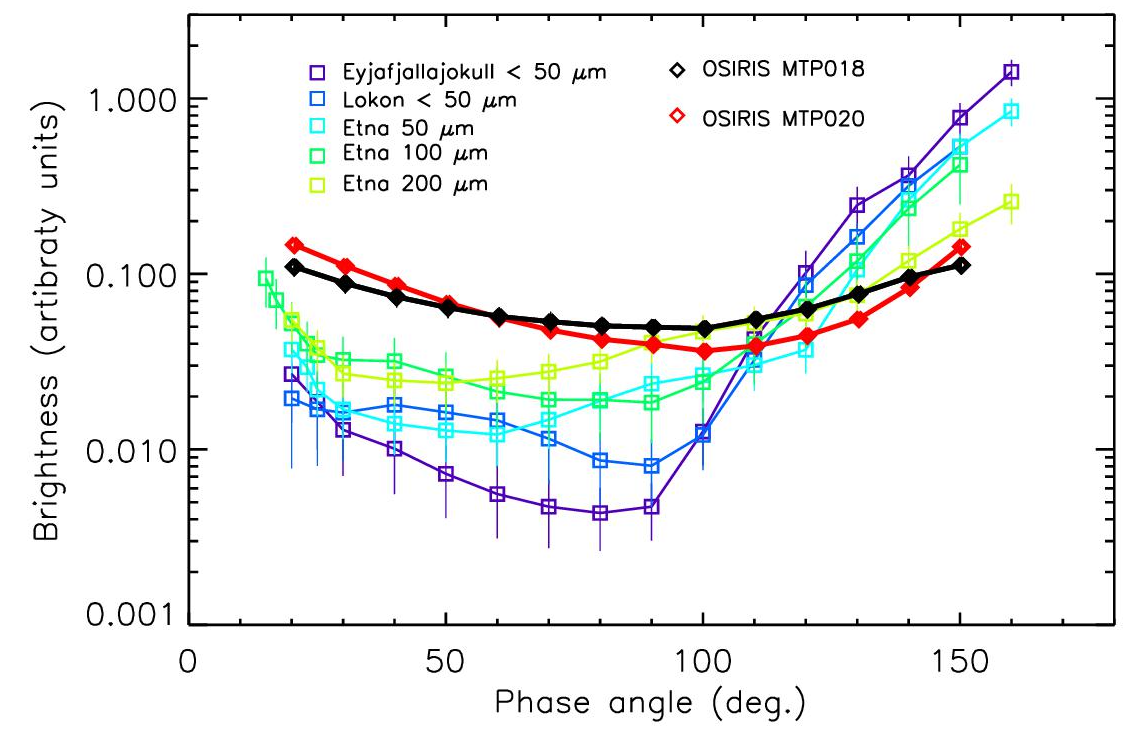}
	\caption{Comparison between OSIRIS and PROGRA2 phase functions
	for porous volcanic particles of different sizes.}
	\label{fig:fig3}
\end{figure}

\begin{figure}
	\center
	\includegraphics[height = 2.2in]{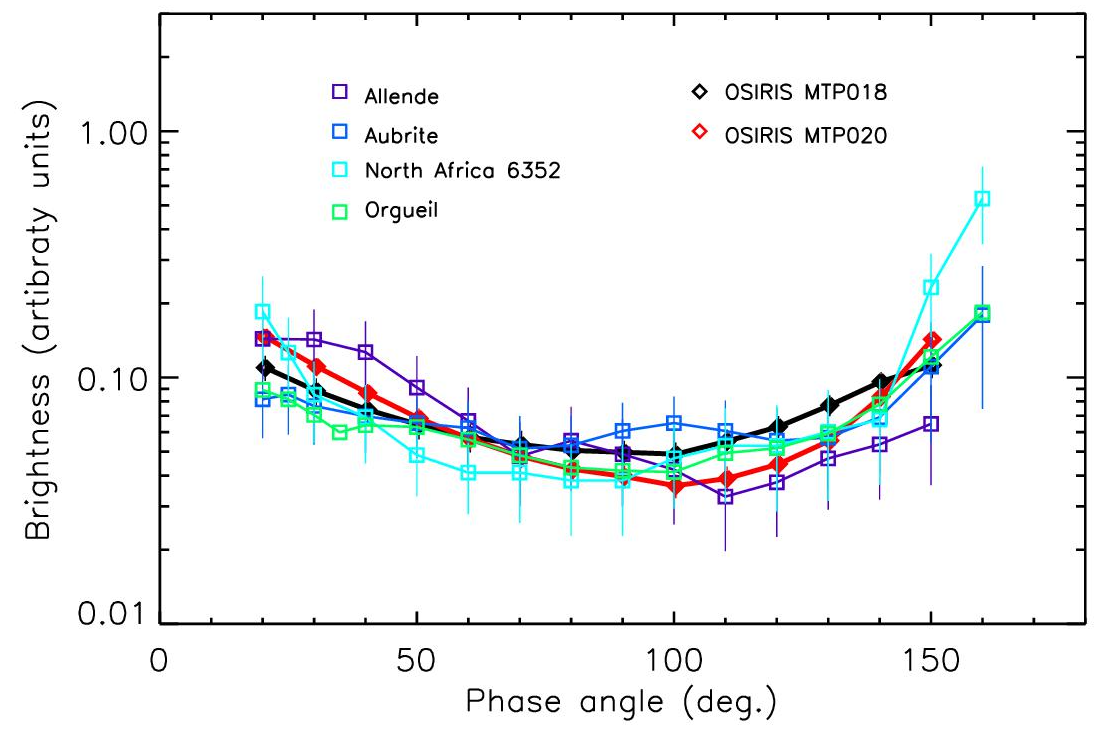}
	\caption{Comparison between OSIRIS and PROGRA2 phase functions 
	for various dust particles from crushed meteorites.}
	\label{fig:fig4}
\end{figure}

\begin{figure}
	\center
	\includegraphics[height = 2.2in]{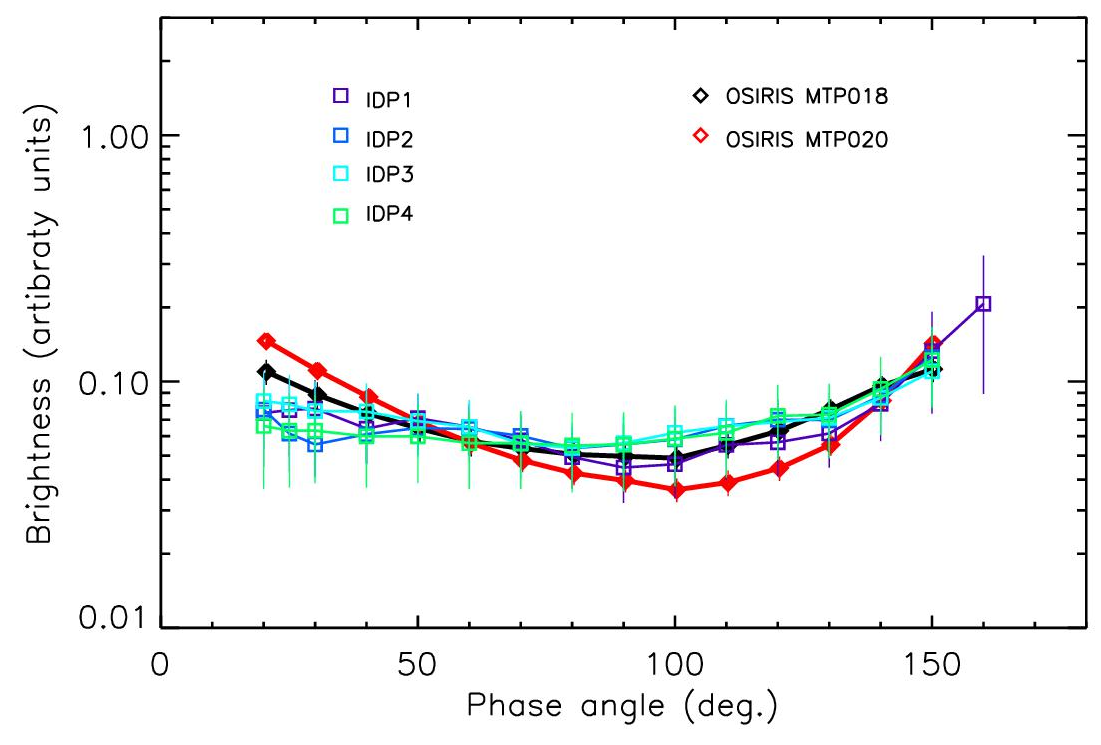}
	\caption{Comparison between OSIRIS and PROGRA2 phase functions
	for various interplanetary dust analogs.}
	\label{fig:fig5}
\end{figure}

  The brightness phase curves measured for the samples described above do not 
  present significant changes with wavelength in the visible domain. 
  Together with their errors bars, they are presented in Figs.~\ref{fig:fig2}, 
  \ref{fig:fig3}, \ref{fig:fig4} and \ref{fig:fig5} 
   for black carbon fluffy aggregates, porous volcanic ashes, 
   crushed meteoric dust particles, and finally for
   interplanetary dust analogs, respectively. 
  The OSIRIS phase curves \citep{bertini2017}, typically obtained in the green 
  domain (by \SI{537}{\nm}) with the wide-angle camera up
  to \ang{150} phase angle, and in 
  the orange domain (by \SI{649}{\nm}) with the narrow-angle
  camera up to \ang{140} phase angle, 
  are also quite comparable. Because the green measurements cover a wider angular 
  coverage, we have chosen to compare them with our laboratory data
   in the green domain. 
  We have considered two sets of data with remarkably small errors bars, 
  MTP018 on 7 July 2015 and MTP020 on 28 August 2015, 
  at heliocentric distances of 1.32 au and 1.25 au, respectively,
  and at distances to the nucleus of 153 km and 420 km, respectively.
  
  It is necessary to normalize the OSIRIS brightness
  phase functions and those from 
  PROGRA2 for comparison purposes. The OSIRIS data have been divided by 
  the sum of the intensities. The PROGRA2 data have also been divided by the sum 
  of the intensity in the \SIrange{20}{150}{\degree} range to
  cover the angular range of OSIRIS. 
  Figures~\ref{fig:fig2} to \ref{fig:fig5} allow immediate comparisons 
  between PROGRA2 and OSIRIS measurements just before and soon after perihelion. 
  Following the types of families described above, they are ordered 
  from poor to better agreement with OSIRIS data.

  The PROGRA2 measurements with black carbon aggregates do not 
  reproduce the OSIRIS phase curves, although the discrepancies 
  decrease with increasing grain sizes. 
  The agreement is also poor for porous volcanic ashes for both 
  small and large phase angles and for the range of phase angles 
  at minimum brightness.
   
  For crushed meteorites, all phase curves present the expected 
  u-shape, with better fits for the crushed carbonaceous chondrites,
  and especially Orgueil, than for the crushed aubrite. 
  Orgueil dust is brownish and is made of porous particles 
  and opaque agglomerates. CO3 6352 is also brownish 
  and is made of irregular relatively porous fragments. 
  Allende fragments are grayer, irregular, and  more compact. 
  Aubrite is clear gray with rather compact fragments. 
  
  Satisfactory agreements are obtained for 
  the interplanetary dust analogs, although their phase curves 
  are relatively flat in the backscattering region. 
  They provide the expected flattened u-shape phase curves.
  
  More precisely, standard deviations are high on average 
   for black carbon ($\approx 0.158$) and for volcanic ashes ($\approx 0.125$). 
  They are significantly lower for crushed meteorites (0.025, with 
  only 0.014 for Orgueil) and for interplanetary dust analogs (0.021).

\section{Discussion}

\subsection{Significance of the different fits}

  The comparison between the brightness phase curves obtained by OSIRIS 
  (before and after perihelion in July-August 2015) and the phase curves 
  obtained by the PROGRA2 instrument on dust samples that might be 
  representative of cometary dust indicates that black carbon fluffy aggregates, 
  porous volcanic ashes, or some crushed meteoritic samples, such as the 
  aubrite and Allende or North Africa 6352 samples  do not provide satisfactory 
  fits. Allende, as also noted in \citet{munoz2000}, 
  presents significant backscattering and a minimum in brightness
  above \ang{90} phase angle.
  

  However, a satisfactory fit is found for a crushed Orgueil sample, which is
  a sample from a carbonaceous meteorite, 
  with a porous structure built of aggregates in a  size range 
  of tens of micrometers \citep{tomeoka1988}.
  A satisfactory fit is also found for samples that have been
  proposed as analogs for 
  interplanetary dust particles.  To our knowledge, 
  they correspond to the first laboratory measurements that
  show a clear u-shaped phase 
  curve, using tentative cometary dust analogs.
  
  The somewhat satisfactory fits are better before than after
  perihelion; the typical standard deviation is 0.008 instead of 0.002 for 
  the Orgueil sample and 0.013 instead 0.022 for the 
  four interplanetary dust analogs. 
  This might correspond to the fact that the population of dust particles 
  that were released after perihelion was less homogeneous than before. 

\subsection{Comparison with clues from the linear polarization of dust analogs }

  Clues from brightness phase curves are, as described in the Introduction, 
  more difficult to interpret than those of linear polarization phase curves, 
  at least when multiple scattering is not significant. Linear polarization phase 
  curves, $P(\alpha)$, usually present a small negative branch in the 
  backscattering region and reach their maximum, $P_{max}$, 
  in the \SIrange{80}{100}{\degree}  phase angle range. They have indeed been 
  extensively used to try to estimate the properties of cometary and 
  interplanetary dust particles, as reviewed in \citet{kiselev2015} and 
  \citet{lasue2015}, respectively. 
  PROGRA2 polarization measurements 
  can provide additional constraints for comparison.
  
  Measurements on three meteoritic dust samples led
  to $P_{max}$ values below 15\%, 
  while the values obtained for the Orgueil sample were higher, about 35\% 
  \citep{hadamcik2018}.
  Previous measurements, made in 1998-1999, with similar 
  Allende and Orgueil samples \citep{worms2000}, have provided similar results. 
  The $P_{max}$ values derived from observations of cometary comae present 
  some dispersion that might be related to cometary activity and to 
  the observational field of view. $P_{max}$ remains
  about 5-10\% for the so-called low-polarization comets,
  and may reach up to 25-30\% for the so-called high-polarization comets. 
  
  Measurements on the four interplanetary dust analogs led to
  $P_{max}$ values in the 20\%-35\% range. While polarimetric phase curves for 
  the zodiacal cloud are quite difficult to obtain because observations along 
  a line of sight correspond to changing phase angles and
  solar distances, 
  inversion is possible in the near-ecliptic symmetry plane of the interplanetary 
  dust cloud \citep{levasseur-regourd2001}. 
  For a phase angle equal to \ang{90}, 
  $P_{max}$ progressively decreases with solar distance from about 35\% at 
  1.5~au to about 28\% at 1~au, and finally about 20\% at 0.4~au.
  This trend, which has long been suspected 
  to originate in partial sublimation of semi-volatile organics,
  is perfectly reproduced 
  with the  interplanetary dust analogs described above, with a decreasing 
  percentage in organics 
  \citep[][Fig. 7]{hadamcik2018}. 
  The agreement between 
  the dust-brightness phase curves of 67P and those of our interplanetary dust 
  analogs is logical, considering that JFCs are now recognized 
  (see the Introduction) to be the main source of interplanetary
  dust along the Earth's orbit.


  Finally, it may be added that although the trends of polarimetric phase curves 
  point out changes in dust properties from the values of the near-maximum 
  polarization (for phase angles in the \ang{90} to \ang{120} range),
  brightness phase 
  curves (when available) may be of major interest for comparisons with 
  simulations results in the back- and forward-scattering regions.
 
\subsection{Comparison with numerical simulations }
  
  Numerical simulations have been developed for the polarimetric phase curves 
  of extensively observed comets, such as 1P/Halley and C/1995 O1 Hale-Bopp. 
  Satisfactory fits to observations have been obtained over a wide range of 
  wavelengths with aggregates of submicron-sized grains mixed with spheroidal 
  particles. Both are consisting of absorbing organic-type material
  and feebly absorbing 
  silicate-type material \citep{levasseur-regourd2008, lasue2009}; 
  interpretations of the linear polarization data for comets are reviewed in 
  \citet{kiselev2015}. 

  
  In relation with brightness phase curves, it was noted in microgravity 
  experiments that u-shaped brightness phase curves for rather spherical 
  glass beads (sizes about \SIrange{50}{150}{\um}) become flattened 
  when the beads are covered with a thin layer of graphite. This  
  shows that carbonaceous compounds may be important
  for the shape of the phase curve in intensity   \citep{lasue2007}. 
  The rougher surface of the carbon-coated beads
  may also have an influence on this effect.
  
  For observations of 67P, \citet{moreno2018} have recently 
  proposed based on numerical simulations that particles with
  sizes above about \SI{20}{\um}, 
  the complex refractive index of which is
  equal to 1.6+0.1i (dark absorbing particles
  at \SI{0.6}{\um}) and the porosity of which is in the 60-70\% range, 
  would fit the OSIRIS phase functions \citep{bertini2017}, assuming that
  the particles are elongated (with long axes perpendicular
  to the solar-radiation 
  direction) and that they have various aspect ratios. 
  Moreover, \citet{markkanen2018} performed 
  numerical simulations that also 
  reproduced the OSIRIS data  with irregular particles,
  with sizes ranging from 
  \SIrange{5}{100}{\um}, that were composed of an intimate
  mixture of sub-micrometer organic 
  material and micrometer-sized spherical silicate grains.
  
  This means that \citet{moreno2018} and \citet{markkanen2018}, although they 
  investigated the free parameter space in their simulations in different ways, 
  both indicate that the phase curves obtained from OSIRIS data in
  the inner coma 
  \citep{bertini2017} can be representative of dust particles
  that are larger than a few 
  microns or tens of microns, and that the absorbing component
  correspond to a very important part. This agrees with our
  laboratory results, where 
  satisfactory fits with the observations are provided by
  irregular absorbing particles 
  (e.g., tens of micrometer-sized crushed carbonaceous meteorites such 
  as Orgueil and interplanetary dust analogs).

\section{Conclusions}

  The dust-brightness phase curves obtained by the OSIRIS observations during 
  the Rosetta rendezvous have revealed unique trends that can be compared with 
  those measured on samples that might be representative of the properties of 
  cometary dust particles. Light-scattering measurements obtained with the 
  PROGRA2 instrument, either in the laboratory or in microgravity conditions, 
  indicate that neither black carbon aggregates, built of grains in the
  range of 14 to \SI{90}{\nm}, 
  nor porous volcanic ashes of various origins and sizes 
  can fairly reproduce the flattened u-shape of the observational phase curves. 
   
  Excellent agreement is nevertheless obtained between the OSIRIS 
  phase curves in July-August 2015 (and especially a few weeks before 
  perihelion) and those measured either on a dust sample from Orgueil 
  meteorite or on samples recognized to be satisfactory analogs for 
  the bulk polarimetric scattering properties of interplanetary dust. 
  The typical shape of the dust-brightness phase curves obtained 
  within the coma of 67P/Churyumov-Gerasimenko is related to a
  significant amount of organic compounds 
  (at least 50\% in mass) and to fluffy aggregates with sizes 
  ranging from 10 to \SI{200}{\um}. These results may be slightly different 
  after perihelion, possibly because dust 
  particles of different sizes and structures were then
  released into the coma. The results nevertheless point out 
  similarities between the dust particles of a Jupiter-family comet and 
  those of dust particles within the inner zodiacal cloud.

\begin{acknowledgements}
ACLR and JL acknowledge support from Centre National d’Études Spatiales (CNES) 
in the scientific analysis of instruments devoted to space exploration of comets. 
The PROGRA2 experiment was funded by CNES; the parabolic flights campaigns 
were funded by CNES and the European Space Agency. 
S. Boresztajn and F. Piller (LISE/CNRS/Sorbonne Univ.) are gratefully 
acknowledged for their SEM images.
\end{acknowledgements}

%
%

\bibliographystyle{aa} 
\bibliography{ACLR_AA2018_lang} 

\begin{thebibliography}{42}
\expandafter\ifx\csname natexlab\endcsname\relax\def\natexlab#1{#1}\fi

\bibitem[{Bardyn {et~al.}(2017)Bardyn, Baklouti, Cottin, Fray, Briois,
  Paquette, Stenzel, Engrand, Fischer, \& Hornung}]{bardyn2017}
Bardyn, A., Baklouti, D., Cottin, H., {et~al.} 2017, Monthly Notices of the
  Royal Astronomical Society, 469, S712

\bibitem[{Bentley {et~al.}(2016)Bentley, Schmied, Mannel, Torkar, Jeszenszky,
  Romstedt, Levasseur-Regourd, Weber, Jessberger, \& Ehrenfreund}]{bentley2016}
Bentley, M.~S., Schmied, R., Mannel, T., {et~al.} 2016, Nature, 537, 73

\bibitem[{Bertini {et~al.}(2017)Bertini, La~Forgia, Tubiana, Güttler, Fulle,
  Moreno, Frattin, Kovacs, Pajola, \& Sierks}]{bertini2017}
Bertini, I., La~Forgia, F., Tubiana, C., {et~al.} 2017, Monthly Notices of the
  Royal Astronomical Society, 469, S404

\bibitem[{Burchell {et~al.}(2008)Burchell, Fairey, Wozniakiewicz, Brownlee,
  Hörz, Kearsley, See, Tsou, Westphal, \& Green}]{burchell2008}
Burchell, M.~J., Fairey, S.~A., Wozniakiewicz, P., {et~al.} 2008, Meteoritics
  \& Planetary Science, 43, 23

\bibitem[{Carrillo-S\'anchez {et~al.}(2016)Carrillo-S\'anchez, Nesvorn\`y,
  Pokorn\`y, Janches, \& Plane}]{carrillo-sanchez2016}
Carrillo-S\'anchez, J.~D., Nesvorn\`y, D., Pokorn\`y, P., Janches, D., \&
  Plane, J. M.~C. 2016, Geophysical research letters, 43

\bibitem[{Dollfus {et~al.}(1988)Dollfus, Bastien, Le~Borgne, Levasseur-Regourd,
  \& Mukai}]{dollfus1988}
Dollfus, A., Bastien, P., Le~Borgne, J.-F., Levasseur-Regourd, A.-C., \& Mukai,
  T. 1988, Astronomy and Astrophysics, 206, 348

\bibitem[{Fink \& Doose(2018)}]{fink2018}
Fink, U. \& Doose, L. 2018, Icarus, 309, 265

\bibitem[{Francis {et~al.}(2011)Francis, Renard, Hadamcik, Couté, Gaubicher,
  \& Jeannot}]{francis2011}
Francis, M., Renard, J.-B., Hadamcik, E., {et~al.} 2011, Journal of
  Quantitative Spectroscopy and Radiative Transfer, 112, 1766

\bibitem[{Fray {et~al.}(2016)Fray, Bardyn, Cottin, Altwegg, Baklouti, Briois,
  Colangeli, Engrand, Fischer, \& Glasmachers}]{fray2016}
Fray, N., Bardyn, A., Cottin, H., {et~al.} 2016, Nature, 538, 72

\bibitem[{Fray {et~al.}(2017)Fray, Bardyn, Cottin, Baklouti, Briois, Engrand,
  Fischer, Hornung, Isnard, \& Langevin}]{fray2017}
Fray, N., Bardyn, A., Cottin, H., {et~al.} 2017, Monthly Notices of the Royal
  Astronomical Society, 469, S506

\bibitem[{Fulle {et~al.}(2018)Fulle, Bertini, Della~Corte, Güttler, Ivanovski,
  La~Forgia, Lasue, Levasseur-Regourd, Marzari, \& Moreno}]{fulle2018}
Fulle, M., Bertini, I., Della~Corte, V., {et~al.} 2018, Monthly Notices of the
  Royal Astronomical Society, 476, 2835

\bibitem[{Fulle \& Blum(2017)}]{fulle2017}
Fulle, M. \& Blum, J. 2017, Monthly Notices of the Royal Astronomical Society,
  469, S39

\bibitem[{Fulle {et~al.}(2000)Fulle, Levasseur-Regourd, McBride, \&
  Hadamcik}]{fulle2000}
Fulle, M., Levasseur-Regourd, A.~C., McBride, N., \& Hadamcik, E. 2000, The
  Astronomical Journal, 119, 1968

\bibitem[{Glassmeier {et~al.}(2007)Glassmeier, Boehnhardt, Koschny, Kührt, \&
  Richter}]{glassmeier2007}
Glassmeier, K.-H., Boehnhardt, H., Koschny, D., Kührt, E., \& Richter, I.
  2007, Space Science Reviews, 128, 1

\bibitem[{Goesmann {et~al.}(2015)Goesmann, Rosenbauer, Bredehöft, Cabane,
  Ehrenfreund, Gautier, Giri, Krüger, Le~Roy, \& MacDermott}]{goesmann2015}
Goesmann, F., Rosenbauer, H., Bredehöft, J.~H., {et~al.} 2015, Science, 349,
  aab0689

\bibitem[{Hadamcik {et~al.}(2018)Hadamcik, Lasue, Levasseur-Regourd, \&
  Renard}]{hadamcik2018}
Hadamcik, E., Lasue, J., Levasseur-Regourd, A.~C., \& Renard, J.-B. 2018,
  Planetary and Space Science

\bibitem[{Hadamcik {et~al.}(2011)Hadamcik, Levasseur-Regourd, Renard, Lasue, \&
  Sen}]{hadamcik2011}
Hadamcik, E., Levasseur-Regourd, A.~C., Renard, J.-B., Lasue, J., \& Sen, A.~K.
  2011, Journal of Quantitative Spectroscopy and Radiative Transfer, 112, 1881

\bibitem[{Hadamcik {et~al.}(2007)Hadamcik, Renard, Lasue, Levasseur-Regourd,
  Blum, \& Schraepler}]{hadamcik2007}
Hadamcik, E., Renard, J.-B., Lasue, J., {et~al.} 2007, Journal of Quantitative
  Spectroscopy and Radiative Transfer, 106, 74

\bibitem[{Hadamcik {et~al.}(2002)Hadamcik, Renard, Worms, Levasseur-Regourd, \&
  Masson}]{hadamcik2002}
Hadamcik, E., Renard, J.~B., Worms, J.~C., Levasseur-Regourd, A.-C., \& Masson,
  M. 2002, Icarus, 155, 497

\bibitem[{Hilchenbach {et~al.}(2016)Hilchenbach, Kissel, Langevin, Briois,
  Von~Hoerner, Koch, Schulz, Silén, Altwegg, \& Colangeli}]{hilchenbach2016}
Hilchenbach, M., Kissel, J., Langevin, Y., {et~al.} 2016, The Astrophysical
  Journal Letters, 816, L32

\bibitem[{Hörz {et~al.}(2006)Hörz, Bastien, Borg, Bradley, Bridges, Brownlee,
  Burchell, Chi, Cintala, \& Dai}]{horz2006}
Hörz, F., Bastien, R., Borg, J., {et~al.} 2006, science, 314, 1716

\bibitem[{Kiselev {et~al.}(2015)Kiselev, Rosenbush, Levasseur-Regourd, \&
  Kolokolova}]{kiselev2015}
Kiselev, N., Rosenbush, V., Levasseur-Regourd, A.-C., \& Kolokolova, L. 2015,
  in Polarimetry of {Stars} and {Planetary} {Systems}, ed. L.~O. Kolokolova,
  J.~Hough, \& A.~C. Levasseur-Regourd (Cambridge University Press), 379--404

\bibitem[{Langevin {et~al.}(2016)Langevin, Hilchenbach, Ligier, Merouane,
  Hornung, Engrand, Schulz, Kissel, Rynö, \& Eng}]{langevin2016}
Langevin, Y., Hilchenbach, M., Ligier, N., {et~al.} 2016, Icarus, 271, 76

\bibitem[{Lasue {et~al.}(2007)Lasue, Levasseur-Regourd, Fray, \&
  Cottin}]{lasue2007}
Lasue, J., Levasseur-Regourd, A.-C., Fray, N., \& Cottin, H. 2007, Astronomy \&
  Astrophysics, 473, 641

\bibitem[{Lasue {et~al.}(2009)Lasue, Levasseur-Regourd, Hadamcik, \&
  Alcouffe}]{lasue2009}
Lasue, J., Levasseur-Regourd, A.~C., Hadamcik, E., \& Alcouffe, G. 2009,
  Icarus, 199, 129

\bibitem[{Lasue {et~al.}(2015)Lasue, Levasseur-Regourd, \&
  Lazarian}]{lasue2015}
Lasue, J., Levasseur-Regourd, A.-C., \& Lazarian, A. 2015, in Polarimetry of
  {Stars} and {Planetary} {Systems}, ed. L.~Kolokolova, J.~Hough, \& A.~C.
  Levasseur-Regourd (Cambridge University Press), 419--436

\bibitem[{Levasseur-Regourd {et~al.}(2018)Levasseur-Regourd, Agarwal, Cottin,
  Engrand, Flynn, Fulle, Gombosi, Langevin, Lasue, \&
  Mannel}]{levasseur-regourd2018}
Levasseur-Regourd, A.-C., Agarwal, J., Cottin, H., {et~al.} 2018, Space science
  reviews, 214, 64

\bibitem[{Levasseur-Regourd {et~al.}(2001)Levasseur-Regourd, Mann, Dumont, \&
  Hanner}]{levasseur-regourd2001}
Levasseur-Regourd, A.~C., Mann, I., Dumont, R., \& Hanner, M.~S. 2001, in
  Interplanetary {Dust}, ed. E.~Grün, B.~A. Gustafson, S.~Dermott, \&
  H.~Fechtig (Springer), 57--94

\bibitem[{Levasseur-Regourd {et~al.}(2015)Levasseur-Regourd, Renard, Shkuratov,
  \& Hadamcik}]{levasseur-regourd2015}
Levasseur-Regourd, A.-C., Renard, J.-B., Shkuratov, Y., \& Hadamcik, E. 2015,
  in Polarimetry of {Stars} and {Planetary} {Systems}, ed. L.~Kolokolova,
  J.~Hough, \& A.~C. Levasseur-Regourd (Cambridge University Press), 62--80

\bibitem[{Levasseur-Regourd {et~al.}(2008)Levasseur-Regourd, Zolensky, \&
  Lasue}]{levasseur-regourd2008}
Levasseur-Regourd, A.-C., Zolensky, M., \& Lasue, J. 2008, Planetary and Space
  Science, 56, 1719

\bibitem[{Mannel {et~al.}(2016)Mannel, Bentley, Schmied, Jeszenszky,
  Levasseur-Regourd, \& Torkar}]{mannel2016}
Mannel, T., Bentley, M.~S., Schmied, R., {et~al.} 2016, Monthly Notices of the
  Royal Astronomical Society, stw2898

\bibitem[{Markkanen {et~al.}(2018)Markkanen, Agarwal, Väisänen, Penttilä, \&
  Muinonen}]{markkanen2018}
Markkanen, J., Agarwal, J., Väisänen, T., Penttilä, A., \& Muinonen, K.
  2018, The Astrophysical Journal Letters, 868, L16

\bibitem[{Matrajt {et~al.}(2008)Matrajt, Ito, Wirick, Messenger, Brownlee,
  Joswiak, Flynn, Sandford, Snead, \& Westphal}]{matrajt2008}
Matrajt, G., Ito, M., Wirick, S., {et~al.} 2008, Meteoritics \& Planetary
  Science, 43, 315

\bibitem[{Moreno {et~al.}(2018)Moreno, Guirado, Muñoz, Bertini, Tubiana,
  Güttler, Fulle, Rotundi, Della~Corte, \& Ivanovski}]{moreno2018}
Moreno, F., Guirado, D., Muñoz, O., {et~al.} 2018, The Astronomical Journal,
  156, 237

\bibitem[{Mu\~noz \& Hovenier(2015)}]{munoz2015}
Mu\~noz, O. \& Hovenier, J.~W. 2015, in Polarimetry of {Stars} and {Planetary}
  {Systems}, ed. L.~Kolokolova, J.~Hough, \& A.~C. Levasseur-Regourd (Cambridge
  University Press), 130--144

\bibitem[{Mu\~noz {et~al.}(2000)Mu\~noz, Volten, De~Haan, Vassen, \&
  Hovenier}]{munoz2000}
Mu\~noz, O., Volten, H., De~Haan, J.~F., Vassen, W., \& Hovenier, J.~W. 2000,
  Astronomy and Astrophysics, 360, 777

\bibitem[{Nesvorn\`y {et~al.}(2010)Nesvorn\`y, Jenniskens, Levison, Bottke,
  Vokrouhlickỳ, \& Gounelle}]{nesvorny2010}
Nesvorn\`y, D., Jenniskens, P., Levison, H.~F., {et~al.} 2010, The
  Astrophysical Journal, 713, 816

\bibitem[{Renard {et~al.}(2002)Renard, Worms, Lemaire, Hadamcik, \&
  Huret}]{renard2002}
Renard, J.-B., Worms, J.-C., Lemaire, T., Hadamcik, E., \& Huret, N. 2002,
  Applied optics, 41, 609

\bibitem[{Rowan-Robinson \& May(2013)}]{rowan-robinson2013}
Rowan-Robinson, M. \& May, B. 2013, Monthly Notices of the Royal Astronomical
  Society, 429, 2894

\bibitem[{Tomeoka \& Buseck(1988)}]{tomeoka1988}
Tomeoka, K. \& Buseck, P.~R. 1988, Geochimica et Cosmochimica Acta, 52, 1627

\bibitem[{Worms {et~al.}(2000)Worms, Renard, Hadamcik, Brun-Huret, \&
  Levasseur-Regourd}]{worms2000}
Worms, J.-C., Renard, J.-B., Hadamcik, E., Brun-Huret, N., \&
  Levasseur-Regourd, A.~C. 2000, Planetary and Space Science, 48, 493

\bibitem[{Worms {et~al.}(1999)Worms, Renard, Hadamcik, Levasseur-Regourd, \&
  Gayet}]{worms1999}
Worms, J.-C., Renard, J.-B., Hadamcik, E., Levasseur-Regourd, A.-C., \& Gayet,
  J.-F. 1999, Icarus, 142, 281

\end{thebibliography}

\end{document}